\documentclass[11pt]{article}
\usepackage{amsmath, amssymb, amsfonts, amsthm}

\newtheorem{theorem}{Theorem}[section]
\newtheorem{proposition}[theorem]{Proposition}
\newtheorem{corollary}[theorem]{Corollary}
\newtheorem{lemma}[theorem]{Lemma}

\newcommand{\rd}{{\rm d}}
\newcommand{\be}{\begin{equation}}
\newcommand{\ee}{\end{equation}}
\newcommand{\bey}{\begin{eqnarray}}
\newcommand{\eey}{\end{eqnarray}}

\newcommand{\eps}{\varepsilon}

\newcommand{\bx}{{\bf x}}

\renewcommand{\a}{\alpha}

\newcommand{\e}{\varepsilon}

\newcommand{\R}{{\mathbb R}}

\newcommand{\bN}{{\mathbb N}}

\newcommand{\tr}{\mbox{Tr}}

\newcommand{\wt}{\widetilde}
\newcommand{\wh}{\widehat}

\newcommand{\const}{\mathrm{const}}

\newcommand{\cA}{{\cal A}}

\newcommand{\cW}{{\cal W}}
\newcommand{\cK}{{\cal K}}
\newcommand{\cH}{{\cal H}}
\newcommand{\cL}{{\cal L}}

\input epsf

\newcommand{\donothing}[1]{}

\begin{document}

\title{Gross-Pitaevskii Equation as the Mean Field Limit \\ of Weakly
Coupled Bosons}
\author{Alexander Elgart${}^1$, L\'aszl\'o Erd\H os${}^2$\thanks{Partially
supported by NSF grant DMS-0200235 and EU-IHP Network ``Analysis
and Quantum'' HPRN-CT-2002-0027. On leave from School of
Mathematics, GeorgiaTech, USA}\;, Benjamin
Schlein${}^1$\thanks{Partially supported by NSF Postdoctoral
Fellowship}\;  and Horng-Tzer Yau${}^1$\thanks{Partially supported
by NSF grant DMS-0307295 and MacArthur Fellowship. On leave
from Courant Institute, New York University} \\
\\
Department of Mathematics, Stanford University\\ Stanford, CA 94305, USA${}^1$ \\ \\
Institute of Mathematics, University of Munich, \\
Theresienstr. 39, D-80333 Munich, Germany${}^2$
\\}

\maketitle

\begin{abstract}
We consider the dynamics of  $N$ boson systems interacting
through a pair potential $N^{-1} V_a(x_i-x_j)$ where
$V_a (x) = a^{-3} V (x/a)$. We denote the solution
to the $N$-particle Schr\"odinger equation by $\psi_{N, t}$.
Recall that the Gross-Pitaevskii (GP) equation is a nonlinear Schr\"odinger
equation and the GP hierarchy
is an infinite BBGKY hierarchy of equations so that if $u_t$ solves the
GP equation, then the family of $k$-particle density matrices  $\{
\otimes_k u_t, k\ge 1 \}$ solves the GP hierarchy. Under the assumption
that $a = N^{-\eps}$ for $0 < \eps < 3/5$, we prove that
as $N\to \infty$ the limit points of the $k$-particle density
matrices of $\psi_{N,t}$ are solutions of the GP hierarchy with the coupling constant
in the nonlinear term of the GP equation given by $\int V(x) dx$.  The
uniqueness of the solutions to this hierarchy remains an open
question.
\end{abstract}

\section{Introduction}
Consider $N$ bosons in a three dimensional
cube $\Lambda$  with the periodic
boundary condition and volume one. The bosons interact via a two body
potential
\[
V_a (x) = \frac{1}{a^3} V (x/a) .
\]
We assume the potential $V$ to be smooth, positive, and with
compact support. The parameter $a$ determines the range and the
strength of the potential: $a$ and $N$ will be coupled so that $a
\to 0$ as $N \to \infty$.  Thus the potential $V_a$ converges to a
Dirac $\delta$-function. The $N$-body Hamiltonian for the $N$
weakly coupled bosons is thus given by
\begin{equation}\label{eq:ham}
H_N= -\sum_{j=1}^N \Delta_j + \frac{1}{N} \sum_{i\neq j} V_a (x_i
- x_j) .
\end{equation}

The density of the bosons in the cube,  $\rho$, is clearly equal to $N$.
Any given particle typically interacts via the potential $V_a$ with $a^3 N$ other
particles. If $a \gg N^{-1/3}$, there are a lot of interactions
among the $N$ bosons. In this case, the ground state of the system
contains, to the leading order in $N$, no correlation among the
particles. Thus the ground state wave function is a product function to the leading
order. In fact, the same conclusion holds as long as $a \gg
N^{-1}$. The correlations among the particles become important in
the leading order only when $a \simeq N^{-1}$.
This choice corresponds to the so called Gross-Pitaevskii scaling
limit, as pointed out by Lieb, Seiringer and Yngvason \cite{LSY1}
(see \cite{LSSY2} for a review).
For this choice of scaling, i.e., $a \simeq N^{-1}$,
we study the dynamics of the Bose gas in \cite{ESY}.
In this paper,  we consider the cases $a = N^{-\eps}$ for $0 < \eps < 3/5$.

In the following we denote by $x$ a general variable in the box
$\Lambda$. On the other hand  $\bx = (x_1 , \dots x_N)$ denotes a
point in $\Lambda^N$. We will also use the notations $\bx_k = (x_1
, \dots x_k) \in \Lambda^k$ and $\bx_{N-k} = (x_{k+1}, \dots x_N)
\in \Lambda^{N-k}$.

The dynamics of the Bose system is governed by the $N$-body
Schr\"odinger Equation
\begin{equation}\label{eq:Schr}
i \partial_t \psi_{N,t} = H_N \psi_{N,t} \, .
\end{equation}
Here the wave function $\psi_{N,t} \in L_s^2 (\Lambda^N)$,
the subspace of $L^2 (\Lambda^N)$ consisting functions
symmetric with respect to permutations of the $N$
particles. More generally we can describe the $N$ body system by
its density matrix $\gamma_{N,t}$. A density matrix is a positive
self-adjoint operator $\gamma$ acting on $L^2_s (\Lambda^N)$, with
$\tr \, \gamma =1$. The density matrix corresponding to the wave
function $\psi_{N,t}$ is given by the orthogonal projection onto
$\psi_{N,t}$, i.e.,  $\gamma_{N,t} = |\psi_{N,t} \rangle \langle
\psi_{N,t} |$. Quantum mechanical states described by orthogonal projections are called pure states.
In general a density matrix is a weighted average of orthogonal
projection (mixed states).
The time evolution of the density matrix $\gamma_{N,t}$ is given
by
\begin{equation}\label{eq:Heis}
i\partial_t \gamma_{N,t} = [ H_N , \gamma_{N,t}]
\end{equation}
which is equivalent to the Schr\"odinger Equation (\ref{eq:Schr}).

It is useful to introduce the marginal distributions corresponding
to the density matrix $\gamma_{N,t}$. For $k=1,\dots, N$, the
$k$-particle marginal $\gamma_{N,t}^{(k)}$ is defined through its
kernel by
\begin{equation}
\gamma_{N,t}^{(k)} (\bx_k ; \bx'_k) = \int \rd \bx_{N-k} \,
\gamma_{N,t} (\bx_k , \bx_{N-k} ; \bx'_k ; \bx_{N-k})
\end{equation}
where $\bx'_k= (x'_1, \dots , x'_k)$ and where $\gamma_{N,t} (\bx
; \bx')$ denotes the kernel of the density matrix $\gamma_{N,t}$.
{F}rom $\tr \,\gamma_{N,t} = 1$, it immediately follows that
\begin{equation}
\tr \gamma_{N,t}^{(k)} = 1
\end{equation}
for every $k=1 , \dots , N$.

Using (\ref{eq:Heis}) (and the symmetry of $\gamma_{N,t}$ with
respect to permutations of the $N$ particles) we find that the
evolution of the marginal distributions of $\gamma_{N,t}$ is
determined by the following hierarchy of $N$ equations, commonly
called the BBGKY Hierarchy:
\begin{equation}\label{eq:BBGKY}
\begin{split}
i\partial_t \gamma_{N,t}^{(k)} &(\bx_k ;\bx'_k) = \sum_{j=1}^k
(-\Delta_{x_j} + \Delta_{x'_j}) \gamma_{N,t}^{(k)} (\bx_k;\bx'_k) \\
&+ \frac{1}{N} \sum_{j \neq \ell}^{k} (V_a (x_j -x_{\ell}) - V_a
(x'_j - x'_{\ell})) \gamma_{N,t}^{(k)} (\bx_k;\bx'_k) \\ &+ (1
-\frac{k}{N}) \sum_{j=1}^k \int \rd x_{k+1} (V_a (x_j - x_{k+1}) -
V_a (x'_j - x_{k+1})) \gamma_{N,t}^{(k+1)}
(\bx_k,x_{k+1};\bx'_k,x_{k+1}).
\end{split}
\end{equation}
Rewriting this hierarchy in integral form yields
\begin{equation}\label{eq:BBGKY2}
\begin{split}
\gamma_{N,t}^{(k)} &(\bx_k;\bx'_k) = \gamma^{(k)}_{N,0}
(\bx_k;\bx'_k) -i \sum_{j=1}^k \int_0^t \rd s (-\Delta_{x_j} +
\Delta_{x'_j}) \gamma_{N,s}^{(k)} (\bx_k;\bx'_k) \\ &- \frac{i}{N}
\sum_{j \neq \ell}^{k} \int_0^t \rd s \, (V_a (x_j -x_{\ell}) -
V_a (x'_j - x'_{\ell})) \gamma_{N,s}^{(k)} (\bx_k; \bx'_k) \\ &-i
(1 -\frac{k}{N}) \sum_{j=1}^k \int_0^t \rd s \, \int \rd x_{k+1}
\, (V_a (x_j - x_{k+1}) - V_a (x'_j - x_{k+1}))
\gamma_{N,s}^{(k+1)} (\bx_k , x_{k+1};\bx'_k, x_{k+1}).
\end{split}
\end{equation}
Letting $a = N^{-\eps}$ and considering the limit $N\to\infty$,
the BBGKY Hierarchy converges formally to the following infinite
hierarchy of equations:
\begin{equation}\label{eq:GPH}
\begin{split}
\gamma_t^{(k)} (\bx_k;\bx'_k) =\; &\gamma_0^{(k)}(\bx_k;\bx'_k) -i
\sum_{j=1}^k \int_0^t \rd s \, (-\Delta_{x_j} + \Delta_{x'_j})
\gamma_{s}^{(k)}
(\bx_k;\bx'_k) \\
&-ib \sum_{j=1}^k \int_0^t \rd s \, \int \rd x_{k+1} (\delta (x_j
- x_{k+1}) - \delta (x'_j - x_{k+1})) \gamma_{s}^{(k+1)} (\bx_k
,x_{k+1};\bx'_k , x_{k+1}),
\end{split}
\end{equation}
where we defined \[ b = \int \rd x V (x) \] (recall that $V_a
(x)=(1/a^3) V(x/a)$). We will call (\ref{eq:GPH}) the infinite
BBGKY hierarchy, or the Gross-Pitaevskii (GP) hierarchy. Note that
(\ref{eq:GPH}) has a factorized solution. The family of marginal
distributions $\tilde{\gamma}_{t}^{(k)} (\bx_k;\bx'_k)=
\prod_{j=1}^k \phi_t (x_j)\overline{\phi_t (x'_j)}$ is a solution
of (\ref{eq:GPH}) if and only if the function $\phi_t$ satisfies
the non-linear Schr\"odinger Equation
\begin{equation}\label{eq:GPE}
i \partial_t \phi_t (y) = -\Delta \phi_t (y) + b |\phi_t (y)|^2
\phi_t (y) \, .
\end{equation}
This is the Gross-Pitaevskii (GP) equation, except that
the coupling constant in front of the non-linear interaction
is given by $b$. In the standard GP equation, the coupling constant
is  $8\pi a_0$ where $a_0$ is the
scattering length of the unscaled potential $V(x)$.

The aim of this paper is to prove the convergence of solutions of
(\ref{eq:BBGKY}) to solutions of (\ref{eq:GPH}): more precisely we
will prove that the sequence $\Gamma_{N,t} = \{ \gamma_{N,t}^{(k)}
\}_{k=1}^N$ has at least one limit point $\Gamma_{\infty,t} = \{
\gamma_{\infty,t}^{(k)} \}_{k \geq 1}$ with respect to some weak
topology, and that any weak limit point $\Gamma_{\infty,t}$
satisfies the infinite hierarchy (\ref{eq:GPH}). For dimension
$d=1$, the convergence to the GP hierarchy \eqref{eq:GPH} for the
delta potential was established by Adami, Bardos, Golse and Teta
in \cite{ABGT}. If the potential has a weaker singularity, the
convergence to the infinite BBGKY hierarchy was proved in
\cite{BGM}.

In \cite{ESY} we proved the convergence to the GP hierarchy with
the coupling constant $8\pi a_0$ when $a \simeq N^{-1}$. Together
with the result of the present paper, this means that the coupling
constant changes from $8\pi a_0$ to $b$ when $a$ changes from
$N^{-1}$ to $N^{-\eps}$, for $\eps < 3/5$. In fact, the coupling
constant should always be given by $b$, as long as $a =
N^{-\eps}$, with $\eps < 1$. This fact follows from the
observation that the potential $N^{-1} a^{-3} V (x/a)$ has
scattering length of order $1/N$ for every choice of $a =
N^{-\eps}$, $\eps >0$. Hence the ground state correlations live on
the scale $1/N$, for all $\eps
>0$. As long as $\eps <1$, these correlations cannot affect
the coupling constant because the potential live in the much
larger scale $a = N^{-\eps} \gg N^{-1}$. In other words, since the
two-particles correlation function is of the form $1 - \const
/(N|x|)$, for $|x| > N^{-1}$, we have
\begin{equation}
\int \rd x \frac{1}{a^3} V(x/a) \left( 1 -
\frac{\const}{N|x|}\right) = b  - \const \, \frac{1}{Na}
\end{equation}
which equals $b$ in the limit $N\to \infty$, if $a \gg N^{-1}$.
Thus, $a \simeq N^{-1}$ is the only scaling for which the coupling
constant is given by the scattering length of the unscaled
potential $V$.

We remark that in \cite{ESY}, we need to use a modified version of
the Hamiltonian \eqref{eq:ham}, in which two-body interactions are
removed, whenever too many particles come into a small region. In
the present paper, this assumption is not needed.  Moreover, we
prove a much stronger a-priori estimate on the limiting density
matrices, see \eqref{eq:apriori}.

Since the $\delta$-function cannot be bounded by
the Laplace operator, we are
unable to prove the uniqueness of the solutions of (\ref{eq:GPH}).
Hence we cannot conclude the {\it propagation of chaos}.
The best known result in this direction is \cite{EY},
which covers the case of a Coulomb singularity in the potential
in $d=3$. Previously, uniqueness was proved by Hepp \cite{H} and Spohn \cite{Sp}
for bounded potential; Ginibre and Velo \cite{GV} had a completely different approach
for quasifree states.

\section{The Main Result}\label{sec:main}

Since our main result states properties of limit points of the
sequence $\gamma_{N,t}^{(k)}$ for $N \to \infty$, in order to
formulate it, we need to specify a topology on the space of
density matrices.

Quantum mechanical states of a $k$-boson system can be described
by a density matrices $\gamma^{(k)}$: $\gamma^{(k)}$ is a
positive, trace class operator, with trace normalized to one. We
can also identify $\gamma^{(k)}$ with its kernel and consider it
as a distribution in $L^2 (\Lambda^k \times \Lambda^k)$: in fact,
since $\gamma^{(k)}$ is a positive operator with trace equal to
one, its Hilbert Schmidt norm is also bounded by one, and
\begin{equation}
\| \gamma^{(k)} \|_2 := \int \rd \bx_k \rd \bx'_k \, |\gamma^{(k)}
(\bx_k ; \bx_k')|^2 \leq 1 \, .
\end{equation}
For $\Gamma = \{ \gamma^{(k)} \}_{k \geq 1} \in \oplus_{k \geq 1}
L^2 (\Lambda^k \times \Lambda^k)$ we define the norm \be \| \Gamma
\|_{H_-} := \sum_{k=1}^{\infty} 2^{-k} \| \gamma^{(k)} \|_2, \ee
and we put \[ H_- := \{ \Gamma \in \bigoplus_{k \geq 1} L^2
(\Lambda^k \times \Lambda^k): \| \Gamma \|_{H_-} < \infty \}. \]
Analogously, we define \[ H_+ = \{ \Gamma \in \bigoplus_{k \geq 1}
L^2 (\Lambda^k \times \Lambda^k) : \lim_{k \to \infty} 2^k \|
\gamma^{(k)} \|_2 = 0 \},\] and we equip $H_+$ with the norm
\[ \| \Gamma \|_{H_+} = \sup_{k \geq 1} 2^k \|
\gamma^{(k)} \|_2 . \] The Banach space $H_-$ is the dual space to
$H_+$. A sequence $\Gamma_N = \{ \gamma^{(k)}_N \}_{k \geq 1} \in
H_-$ converges to $\Gamma_{\infty} = \{ \gamma^{(k)}_{\infty}
\}_{k \geq 1} \in H_-$ in the weak$*$ topology if and only if \be
\label{eq:topH-} \lim_{N \to \infty} \sum_{k \geq 1} \int \rd
\bx_k \rd \bx'_k \, J^{(k)} (\bx_k, \bx'_k) \left(\gamma^{(k)}_N
(\bx_k, \bx'_k) - \gamma^{(k)}_{\infty} (\bx_k, \bx'_k) \right) =0
\ee for all $J= \{ J^{(k)} \}_{k \geq 1} \in H_+$ (this is
actually equivalent to convergence for each fixed $k$). We will
denote by $C ([0,T] , H_-)$ the space of functions of $t \in
[0,T]$ with values in $H_-$ which are continuous with respect to
the weak star topology on $H_-$. Since the space $H_+$ is
separable, we can fix a dense countable subset in the unit ball of
$H_+$, denoted by $\{J_i\}_{i \ge 1}$. Define the metric on $H_-$
by
\begin{equation}\label{eq:rho}
\rho (\Gamma, \wt \Gamma) : = \sum_{i=1}^\infty 2^{-i} \Big |
\sum_{k=1}^\infty \int \rd \bx_k \rd \bx'_k \overline{J_i^{(k)}
(\bx_k ; \bx'_k)} \big [ \gamma^{(k)}(\bx_k ; \bx'_k)- \wt
\gamma^{(k)}(\bx_k; \bx'_k) \big ] \Big | \; .
\end{equation}
Then the topology induced by $\rho (.,.)$ and the  weak* topology
are equivalent on the unit ball $B_-$ of $H_-$. We equip $C ([0,T]
, H_-)$  with the metric \be \wh \rho (\Gamma, \wt \Gamma) :=
\sup_{0\le t \le T}  \; \rho\,(\Gamma(t), \wt \Gamma(t))\; .
\label{def:varrho} \ee We are now ready to state our main theorem.

\begin{theorem}\label{thm:main}
Assume the potential $V(x)$ is positive, smooth, and has compact
support, and set $V_a (x) = a^{-3} V (x/a)$. Suppose that $a =
N^{-\eps}$, for some $0 < \eps < 3/5$. Choose an initial density
matrix $\gamma_{N,0}$ such that
\[ \tr \, H^{k}_N \gamma_{N,0} \leq C^k N^k \] for some constant
$C$ and for all $k \geq 1$. Let $\Gamma_{N,0} = \{
\gamma_{N,0}^{(k)} \}_{k=1}^N$ be the family of marginal
distributions corresponding to the initial density matrix
$\gamma_{N,0}$. Fix now $T>0$ and denote by $\Gamma_{N,t} = \{
\gamma_{N,t}^{(k)}\}$, for $t \in [0,T]$, the solution to the
BBGKY Hierarchy (\ref{eq:BBGKY}) corresponding to the initial data
$\Gamma_{N,0}$.
\begin{itemize}
\item[i)] The sequence $\Gamma_{N,t}$ is compact in $C([0,T],
H_-)$ with respect to the metric $\wh \rho$. \item[ii)] Let
$\Gamma_{\infty, t} = \{ \gamma^{(k)}_{\infty,t} \}_{k\geq 1} \in
C([0,T], H_-)$ be any limit point of $\Gamma_{N,t}$ with respect
to the metric $\wh \rho$. Then there is a constant $C$ such that
\begin{equation}\label{eq:apriori}
\tr \, \big| S_1 \dots S_k \, \gamma_{\infty,t}^{(k)} S_k \dots
S_1 \big| \leq C^k
\end{equation}
for every $k \geq 1$. Here we use the notation $S_j = (1
-\Delta_j)^{1/2}$.
\item[iii)] $\Gamma_{\infty,t}$ is non trivial. In particular we have
\begin{equation}
\tr \, \gamma_{\infty,t}^{(k)} = 1
\end{equation}
for every $t \in [0,T]$ and for every $k \geq 1$.
\item[iv)] Assume
$h_r (x) = r^{-3} h(x/r)$ for any $h\in
C_0^\infty(\Lambda)$ with $\int_\Lambda h =1$. Then, for any $k
\geq 1$ and $t \in [0,T]$, the limit
\begin{equation}\label{eq:limrr'}
\begin{split}
\lim_{r,r' \to 0} \int \rd x'_{k+1} \rd x_{k+1} \,
h_r (x'_{k+1} - x_{k+1}) h_{r'} &(x_{k+1} - x_j)
\gamma^{(k)}_{\infty,t} (\bx_k , x_{k+1} ; \bx'_k , x'_{k+1}) \\
&= :
\gamma^{(k)}_{\infty,t} (\bx_k, x_j ; \bx'_k , x_j)
\end{split}
\end{equation}
exists in the weak $W^{-1,1} (\Lambda^k \times \Lambda^k)$-sense
and defines $\gamma^{(k)} (\bx_k, x_j ; \bx'_k , x_j)$ as a
distribution of $2k$ variables \footnote{Here and in the following
$W^{p,q}(\Lambda^k \times \Lambda^k)$ denotes the usual Sobolev
space over $\Lambda^k \times \Lambda^k$.}.
\item[v)]
$\Gamma_{\infty,t}$ satisfies the infinite Gross-Pitaevski
Hierarchy (\ref{eq:GPH}) in the following sense: For any $J^{(k)}
({\bf x}_k,{\bf x}'_k) \in W^{2,\infty} (\Lambda^k \times
\Lambda^k)$ we have
\begin{equation}\label{eq:GPH3}
\begin{split}
\int \rd {\bf x_k} \rd {\bf x}_k' \, J^{(k)} ({\bf x}_k , {\bf
x}_k') \gamma_{\infty,t}^{(k)} &({\bf x}_k , {\bf x}_k') = \int
\rd {\bf x}_k \rd {\bf x}_k' J^{(k)} ({\bf x}_k , {\bf x}_k')
\gamma_{\infty,0}^{(k)} ({\bf x}_k ,{\bf x}_k') \\
-i \sum_{j=1}^k \int_0^t &\rd s \, \int \rd {\bf x}_k \rd {\bf
x}_k' \, J^{(k)} ({\bf x}_k, {\bf x}_k') \, ( -\Delta_j +
\Delta_j') \gamma^{(k)}_{\infty,s} ({\bf x}_k, {\bf x}_k')
\\ - i b \sum_{j=1}^k &\int_0^t \rd s \int \rd {\bf x}_k \rd {\bf x}_k' \rd
x_{k+1} \, J^{(k)} ({\bf x}_k ,{\bf x}_k') (\delta (x_j - x_{k+1})
- \delta (x_j' - x_{k+1}))
\\ &\times \gamma^{(k+1)}_{\infty,s} ({\bf x}_k , x_{k+1}; {\bf
x}_k',x_{k+1})\, .
\end{split}
\end{equation}
Here the action of the $\delta$-functions on $\gamma^{(k+1)}$ is
well defined (through a regularization of the $\delta$-function)
by part iv).
\end{itemize}
\end{theorem}

\section{Energy Estimates}

The main tool in the proof of Theorem \ref{thm:main} is the
following proposition, which proves bounds for the $L^2$-norm of
the derivatives of a wave function $\psi$ in terms of the
expectation of powers of the Hamiltonian $H_N$ (defined in
(\ref{eq:ham})) in the state described by $\psi$. The proof of
this proposition requires some standard Sobolev-type inequalities,
which are collected, for completeness, in Appendix \ref{app:sob}.

\begin{proposition}\label{prop:energy}
Suppose $V$ is smooth and positive. Put $V_a (x) = a^{-3} V (x/a)$
and assume $a=N^{-\eps}$, with $0 < \eps < 3/5$. Put
\[ \tilde{H}_N = \sum_{j=1}^N S_j^2 + \frac{1}{N} \sum_{\ell \neq
m} V_a (x_{\ell} - x_m) = H_N + N .\] Fix $k\in \bN$ and $0 < C
<1$. Then there is $N_0 = N_0 (k,C)$ such that
\begin{equation}
(\psi, \tilde{H}_N^k \psi ) \geq C^k N^k \, (\psi, S_1^2 S_2^2
\dots S_k^2 \psi)
\end{equation}
for all $N > N_0$ and all $\psi \in D(H_N^k)$ ($\psi$ is assumed
to be symmetric with respect to any permutation of all its
variables).
\end{proposition}
\begin{proof}
The proof of the proposition is by a two step induction over $k$.
For $k=0$ and $k=1$ the claim is trivial (because of the
positivity of the potential). Now we assume the proposition is
true for all $k \leq n$, and we prove it for $k=n+2$. To this end
we apply the induction assumption and we find, for $N > N_0
(n,C)$,
\begin{equation}\label{eq:induction}
(\psi, \tilde{H}_N^{n+2} \psi) = (\psi, \tilde{H}_N
\tilde{H}_N^{n} \tilde{H}_N \psi) \geq C^n N^n (\psi \tilde{H}_N
S_1^2 \dots S_n^2 \tilde{H}_N \psi )\, .
\end{equation}
We put \[ H^{(n)} = \sum_{j=1}^n S_j^2 + \frac{1}{N} \sum_{j <
m}^N V_{jm} \] with $V_{jm} = a^{-3} V((x_j -x_m)/a)$. Then we
have
\begin{equation*}
\begin{split}
(\psi \tilde{H}_N S_1^2 \dots S_n^2 \tilde{H}_N \psi ) = \;
&\sum_{j_1, j_2 \geq n+1} (\psi, S_{j_1}^2 S_1^2 \dots S_n^2
S_{j_2}^2 \psi) + \sum_{j \geq n+1} \left((\psi , S_j^2 S_1^2
\dots S_n^2 H^{(n)} \psi) + \text{c.c.} \right) \\ &+ (\psi ,
H^{(n)} S_1^2 \dots S_n^2 H^{(n)} \psi) \, .
\end{split}
\end{equation*}
Since $H^{(n)} S_1^2 \dots S_n^2 H^{(n)} \geq 0$, we find, using
the symmetry with respect to permutations,
\begin{equation}\label{eq:ener1}
\begin{split}
(\psi \tilde{H}_N S_1^2 \dots S_n^2 \tilde{H}_N \psi ) \geq &\;
(N-n) (N-n-1) (\psi, S_1^2 \dots S_{n+2}^2 \psi) + (2n +1) (N-n)
(\psi, S_1^4 S_2^2 \dots S_{n+1}^2 \psi) \\ &+
\frac{n(n+1)(N-n)}{2N} \left( (\psi, V_{12} S_1^2 \dots S_{n+1}^2
\psi) + \text{c.c.} \right) \\ &+ \frac{(n+1)(N-n)(N-n-1)}{N}
\left( (\psi, V_{1,n+2} S_1^2 \dots S_{n+1}^2 \psi) + \text{c.c.}
\right) \, ,
\end{split}
\end{equation}
where c.c. denotes the complex conjugate. Here we also used that $
(\psi , V_{jm} S_1^2 \dots S_{n+1}^2 \psi) \geq 0 $ if $j,m >
n+1$, because of the positivity of the potential. Next we consider
the term on the second line of (\ref{eq:ener1}): note that this
term vanishes if $n=0$, so we can assume $n \geq 1$. Then we
have
\begin{equation*}
\begin{split}
(\psi, V_{12} S_1^2 &\dots S_{n+1}^2 \psi) + \text{c.c.} = \;
(\psi, S_{n+1} \dots S_3 V_{12} S_1^2 S_2^2 S_3 \dots S_{n+1}
\psi) + \text{c.c.} \\ = \;& (\psi, S_{n+1} \dots S_3 V_{12}
(1+p_1^2)(1+p_2^2) S_3 \dots S_{n+1} \psi) + \text{c.c.} \\ \geq
\; &2 (\psi, S_{n+1} \dots S_3 V_{12} p_2^2 S_3 \dots S_{n+1}
\psi) + (\psi , S_{n+1} \dots S_3 V_{12} p_1^2 p_2^2 S_3 \dots
S_{n+1} \psi) + \text{c.c.} \\ \geq \;& 2 (\psi, S_{n+1} \dots S_3
\nabla V_{12} p_2 S_3 \dots S_{n+1} \psi)  + (\psi , S_{n+1} \dots
S_3 p_2 \nabla V_{12} p_1 p_2 S_3 \dots S_{n+1} \psi)
\\ &+ (\psi, S_{n+1} \dots S_3 \nabla V_{12} p_1^2 p_2 S_3 \dots
S_{n+1} \psi) + \text{c.c.}
\end{split}
\end{equation*}
where $\nabla V_{12} = a^{-4} (\nabla V) ((x_1 -x_2)/a)$. Applying
Schwarz inequality we get
\begin{equation*}
\begin{split}
(\psi, &V_{12} S_1^2 \dots S_{n+1}^2 \psi) + \text{c.c.}  \\
\geq \; &- 2 \{ \a_1 (\psi, S_{n+1} \dots S_3 |\nabla V_{12}| S_3
\dots S_{n+1} \psi) + \a_1^{-1} (\psi, S_{n+1} \dots S_3 |p_2|
|\nabla V_{12}| |p_2|
S_3 \dots S_{n+1} \psi) \} \\
& - \{ \a_2 (\psi , S_{n+1} \dots S_3 |p_2| |\nabla V_{12}| |p_2|
S_3 \dots S_{n+1} \psi)  \\ &\hspace{5cm} + \a_2^{-1} (\psi ,
S_{n+1} \dots S_3 |p_2| |p_1|
|\nabla V_{12}| |p_1| |p_2| S_3 \dots S_{n+1} \psi) \}  \\
&- \{ \a_3 (\psi, S_{n+1} \dots S_3 |\nabla V_{12}| S_3 \dots
S_{n+1} \psi) + \a_3^{-1} (\psi , S_{n+1} \dots S_3 |p_2| p_1^2
|\nabla V_{12}| p_1^2 |p_2| S_3 \dots S_{n+1} \psi) \} \,
.\end{split}
\end{equation*}
Using Lemma \ref{lm:sob} we find
\begin{equation}\label{eq:ener2}
\begin{split}
(\psi, V_{12} &S_1^2 \dots S_{n+1}^2 \psi) + \text{c.c.} \\ \geq
\;&- C \{ \a_1 a^{-1} (\psi, S_{n+1} \dots S_3 S_1^2 S_2^2
S_3\dots S_{n+1}\psi ) + \a_1^{-1} a^{-2} (\psi , S_ {n+1}
\dots S_3 S_1^2 p_2^2 S_3 \dots S_{n+1} \psi) \} \\
&- C \{ \a_2 a^{-2} (\psi, S_{n+1} \dots S_3 S_1^2 p_2^2 S_3\dots
S_{n+1}\psi ) + \a_2^{-1} a^{-2} (\psi , S_
{n+1} \dots S_3 S_1^2 p_1^2 p_2^2 S_3 \dots S_{n+1} \psi) \} \\
&- C \{  \a_3 a^{-1} (\psi, S_{n+1} \dots S_3 S_1^2 S_2^2 S_3
\dots S_{n+1}\psi) + \a_3^{-1} a^{-4}  (\psi , S_ {n+1} \dots S_3
p_2^2 p_1^4 S_3 \dots S_{n+1} \psi) \}
\\ \geq \; &- C N^{-3/2} a^{-5/2} \{ N^2 (\psi, S_1^2 S_2^2 S_3^2 \dots
S_{n+1}^2 \psi ) + N (\psi , S_1^4 S_2^2 \dots S_{n+1}^2 \psi) \}
\end{split}
\end{equation}
where we optimized the choice of $\a_1, \a_2,\a_3$. As for the
last term on the r.h.s. of (\ref{eq:ener1}) we have
\begin{equation*}
\begin{split}
(\psi, V_{1,n+2} S_1^2 &\dots S_{n+1}^2 \psi) + \text{c.c.} =
(\psi, S_{n+1} \dots S_2 V_{1, n+2} S_1^2 S_2 \dots S_{n+1} \psi)
+ \text{c.c.} \\ \geq \; & (\psi, S_{n+1} \dots S_2 V_{1, n+2}
p_1^2 S_2 \dots S_{n+1} \psi) + \text{c.c.} \\ \geq \; &(\psi,
S_{n+1} \dots S_2 \nabla V_{1,n+2} p_1 S_2 \dots S_{n+1} \psi) +
\text{c.c.} \\ \geq \;&-\a (\psi, S_{n+1} \dots S_2 |\nabla
V_{1,n+2}| S_2 \dots S_{n+1} \psi) \\ &\hspace{5cm} - \a^{-1}
(\psi, S_{n+1} \dots
S_2 |p_1| |\nabla V_{1,n+2}| |p_1| S_2 \dots S_{n+1} \psi) \\
\geq \;& - C (\a a^{-1}+ \a^{-1} a^{-2}) \, (\psi, S_1^2 \dots
S_{n+2}^2 \psi) \\ \geq \; & - C a^{-3/2} (\psi, S_1^2 \dots
S_{n+2}^2 \psi)\, .
\end{split}
\end{equation*}
Inserting last equation and (\ref{eq:ener2}) in the r.h.s. of
(\ref{eq:ener1}) we get
\begin{equation*}
\begin{split}
(\psi \tilde{H}_N S_1^2 \dots S_n^2 \tilde{H}_N \psi ) \geq &\;
(N-n) (N-n-1) \left(1 - \frac{C}{N a^{3/2}}
-\frac{C}{N^{3/2}a^{5/2}}\right) (\psi, S_1^2 \dots S_{n+2}^2
\psi)
 \\ &+ (2n +1) (N-n) \left(1 - \frac{C}{N^{3/2}a^{5/2}}\right) (\psi, S_1^4
S_2^2 \dots S_{n+1}^2 \psi) \, .
\end{split}
\end{equation*}
Since $a = N^{-\e}$ with $\e < 3/5$, we have $N^{3/2} a^{5/2} \gg
1$ and $N a^{3/2} \gg 1$. {F}rom the last equation, for any fixed
$C <1$ and $n \in \bN$, we can find $N_0$ so that
\begin{equation}
(\psi ,\tilde{H}_N S_1^2 \dots S_n^2 \tilde{H}_N \psi ) \geq C^2
N^2 (\psi, S_1^2 \dots S_{n+2}^2 \psi)
\end{equation}
for every $N \geq N_0$. This, together with (\ref{eq:induction})
completes the proof of the proposition.
\end{proof}

\begin{corollary}\label{cor:energy}
Suppose the initial density matrix $\gamma_{N,0}$ satisfies \[ \tr
\, H_N^k \gamma_{N,0} \leq C_1^k N^k \, .\] Let $\gamma_{N,t}$ be
the solution of \eqref{eq:Heis}, and let $\{ \gamma^{(k)}_{N,t}
\}_{k=0}^N$ be the corresponding marginal distributions. Then, for
any $C \geq C_1$ and any $k \in \bN$ there is $N_0 = N_0 (k,C)$
such that
\begin{equation}
\tr \, S_1 \dots S_{k} \gamma^{(k)}_{N,t} S_k \dots S_1 \leq C^k
\end{equation}
for all $t \in \R$ and all $N \geq N_0$.
\end{corollary}


\section{Proof of the Main Theorem}

In this section we prove our main result, Theorem \ref{thm:main}.
To this end we will make use of the following lemma. We use here
the notation
\begin{equation}
\langle J^{(k)} , \gamma^{(k)} \rangle = \int \rd \bx_k \rd \bx'_k
J^{(k)} (\bx_k,\bx'_k) \, \gamma^{(k)} (\bx_k,\bx'_k) \, .
\end{equation}

\begin{lemma}\label{lm:conv}
Fix $k \geq 1$ and $J^{(k)} \in W^{1,\infty} (\Lambda^k \times
\Lambda^k)$. For $\beta
>0$ set $\delta_{\beta} (x) = (4\pi \beta^3 /3)^{-1} \chi ( |x|
\leq \beta)$. Then we have
\begin{equation}
\begin{split}
\langle J^{(k)} , &\gamma^{(k)}_{N,t} \rangle = \langle J^{(k)} ,
\gamma^{(k)}_{N,0} \rangle -i \sum_{j=1}^k \int_0^t \rd s \int \rd
\bx_k \rd \bx'_k J^{(k)} (\bx_k,\bx'_k) (-\Delta_{x_j} +
\Delta_{x'_j}) \gamma_{N,s}^{(k)} (\bx_k, \bx'_k) \\ & -ib
\sum_{j=1}^k \int_0^t \rd s \int \rd \bx_k \rd \bx'_k \rd x_{k+1}
J^{(k)} (\bx_k,\bx'_k) (\delta_{\beta} (x_j - x_{k+1}) -
\delta_{\beta} (x'_j - x_{k+1})) \\
&\hspace{3cm} \times \gamma_{N,s}^{(k+1)} (\bx_k,x_{k+1}, \bx'_k,
x_{k+1})\\&+ t \left( k O(\beta^{1/2}) + k O(a^{1/2}) + k^2 O
\left( \frac{1}{Na^{1/2}}\right) \right) \, \sup_{s \in [0,t]} \tr
|S_1 S_2 \gamma^{(2)}_{N,s} S_2 S_1| \, .
\end{split}
\end{equation}
\end{lemma}

\begin{proof} We start from the BBGKY Hierarchy
(\ref{eq:BBGKY2}). After multiplying with $J^{(k)} \in L^2
(\Lambda^k \times \Lambda^k)$ we get
\begin{equation}
\begin{split}
\langle J^{(k)} , \gamma_{N,t}^{(k)} \rangle = \; &\langle J^{(k)}
, \gamma^{(k)}_{N,0} \rangle -i \sum_{j=1}^k \int_0^t \rd s \int
\rd \bx_k \rd \bx'_k J^{(k)} (\bx_k,\bx'_k) (-\Delta_{x_j}
+ \Delta_{x'_j}) \gamma_{N,s}^{(k)} (\bx_k, \bx'_k) \\
&- \frac{i}{N} \sum_{j \neq \ell}^{k} \int_0^t \rd s \, \int \rd
\bx_k \rd \bx'_k \, J^{(k)} (\bx_k,\bx'_k) (V_a (x_j -x_{\ell}) -
V_a (x'_j - x'_{\ell}))
\gamma_{N,s}^{(k)} (\bx_k,\bx'_k) \\
&-i (1 -\frac{k}{N}) \sum_{j=1}^k \int_0^t \rd s \, \int \rd \bx_k
\rd \bx'_k \rd x_{k+1} \, J^{(k)} (\bx_k,\bx'_k) \, (V_a (x_j -
x_{k+1}) - V_a (x'_j - x_{k+1})) \\ &\hspace{4cm} \times
\gamma_{N,s}^{(k+1)} (\bx_k,x_{k+1},\bx'_k,x_{k+1}).
\end{split}
\end{equation}
Next we estimate
\begin{equation}
\begin{split}
\Big| \int \rd \bx_k \rd \bx'_k \, J^{(k)} (\bx_k,\bx'_k) &V_a
(x_j -x_{\ell}) \gamma_{N,s}^{(k)} (\bx_k,\bx'_k) \Big| = \Big|
\tr J^{(k)} V_a (x_j -x_{\ell}) \gamma_{N,s}^{(k)} \Big|
\\ &\leq \| S_j^{-1} J^{(k)} S_j \| \, \| S_j^{-1} V_a (x_j
-x_{\ell}) S_j^{-1} S_{\ell}^{-1} \| \, \| S_{\ell}^{-1} \| \, \tr
S_j S_{\ell} \gamma^{(k)}_{N,s} S_j S_{\ell} \\ &\leq C_k a^{-1/2}
\, \tr \, S_1 S_2 \gamma^{(2)}_{N,s} S_2 S_1
\end{split}
\end{equation}
where we used that, by Lemma \ref{lm:sob}, $\| (V_a)^{1/2} (x_j
-x_{\ell})  S_j^{-1} \| \leq C a^{-1/2}$, and $\| (V_a)^{1/2} (x_j
-x_{\ell})  S_j^{-1} S_{\ell}^{-1} \| \leq C$.  Moreover we used
that $\| S_j^{-1} J^{(k)} S_j \| \leq \| J^{(k)} S_j \|$ and that
\begin{equation}\label{eq:SJ}
\| J^{(k)} S_j \|^2 \leq \| J^{(k)} S_j^2 (J^{(k)})^* \| \leq \tr
\, J^{(k)} S_j^2 (J^{(k)})^*  = \int \rd \bx_k \rd \bx'_k \,
\left(|J^{(k)}(\bx_k , \bx'_k)|^2 + |\nabla_j J^{(k)} (\bx_k,
\bx'_k)|^2 \right)
\end{equation}
which is bounded, because of the finiteness of the volume, and
because, by assumption $J^{(k)} \in W^{1,\infty} (\Lambda^k \times
\Lambda^k)$.

In the same way we can bound the contribution arising from the
term $V_a (x'_j - x'_{\ell})$, and so we find
\begin{multline}
\Big| \frac{1}{N} \sum_{j \neq \ell}^{k} \int_0^t \rd s \, \int
\rd \bx_k \rd \bx'_k \, J^{(k)} (\bx_k,\bx'_k) (V_a (x_j
-x_{\ell}) - V_a (x'_j - x'_{\ell})) \gamma_{N,s}^{(k)}
(\bx_k,\bx'_k) \Big| \\ \leq  \frac{ C_k t}{Na^{1/2}} \, \sup_{s
\in [0,t]} \, \tr \, S_1 S_2 \gamma^{(2)}_{N,s} S_2 S_1 \, .
\end{multline}
Analogously we also get
\begin{equation}
\begin{split}
\Big|\frac{k}{N} \sum_{j=1}^k \int_0^t \rd s &\, \int \rd \bx_k
\rd \bx'_k \rd x_{k+1} \, J^{(k)} (\bx_k,\bx'_k) \, (V_a (x_j -
x_{k+1}) - V_a (x'_j -
x_{k+1})) \\ & \times\gamma_{N,s}^{(k+1)} (\bx_k,x_{k+1},\bx'_k,x_{k+1})\Big| \\
\leq \; &\frac{C_k t}{N} \|J^{(k)} \| \| S_{k+1}^{-1} V_a (x_j -
x_{k+1}) S_{k+1}^{-1} S_j^{-1} \| \| S_j^{-1} \|  \sup_{s \in
[0,t]} \tr \, S_1 S_2 \gamma^{(2)}_{N,s} S_2 S_1  \\
\leq \; &\frac{C_k t}{Na^{1/2}} \sup_{s \in [0,t]} \tr \, S_1 S_2
\gamma^{(2)}_{N,s} S_2 S_1 \, .
\end{split}
\end{equation}
Applying Lemma \ref{lm:sobsob} twice (once with $\beta_2 =a$ and
once with $\beta_2 = \beta$; in both cases with $\beta_1 =0$), we
have
\begin{multline*}
\Big| \int \rd \bx_k \rd \bx'_k \, J^{(k)} (\bx_k, \bx'_k) \left(
V_a (x_j -x_{k+1}) - b \delta_{\beta} (x_j -x_{k+1})\right)
\gamma^{(k+1)}_{N,s} (\bx_k,x_{k+1},\bx'_k , x_{k+1}) \Big| \\
\leq C (a^{1/2} + \beta^{1/2}) \, \tr \, S_1 S_2
\gamma^{(2)}_{N,s} S_2 S_1
\end{multline*}
for some constant $C$ which only depends on $J^{(k)}$, but is
independent of $N$, of $\beta$, and of $s \in [0,t]$.
\end{proof}

The following lemma is used to regularize the action of the
$\delta$-function. It was already used in the proof of Lemma
\ref{lm:conv}. It will be used again to prove the convergence to
the infinite BBGKY hierarchy. Its proof can be found in \cite{ESY}
(see Proposition 8.1).

\begin{lemma}\label{lm:sobsob}
Suppose $\delta_{\beta} (x)$ is a radially symmetric function,
with $0 \leq \delta_{\beta} (x) \leq C \beta^{-3} \chi (|x| \leq
\beta)$ and $\int \delta_{\beta} (x) \rd x = 1$ (for example
$\delta_{\beta} (x) = \beta^{-3} h (x/\beta)$, for a radially
symmetric probability density $h(x)$ supported in $\{ x : |x| \leq
1\}$). Then, for any $J^{(k)} \in W^{1,\infty} (\Lambda^k \times
\Lambda^k)$ and for any smooth function $\gamma^{(k+1)} (\bx_{k+1}
; \bx'_{k+1})$ corresponding to a $(k+1)$-particle density matrix,
we have, for any fixed $j \leq k$,
\begin{equation}\label{eq:Uintbound}
\begin{split}
\Big| \int \rd {\bf x}_k \rd {\bf x}_k' \rd x_{k+1} \rd x'_{k+1}
\, J^{(k)} (\bx_k ; &\bx'_k) \, \left(\delta_{\beta_1} (x'_{k+1} -
x_{k+1}) \delta_{\beta_2} (x_j -x_{k+1}) - \delta (x'_{k+1} -
x_{k+1}) \delta (x_j - x_{k+1})\right) \\ \times \gamma^{(k+1)}
(\bx_k, x_{k+1}; &\bx'_k , x'_{k+1}) \Big|  \\ \leq \; &C [\,
\|J\|_\infty + \|\nabla_j J \|_\infty \, \big ] \, (\beta_1 +
\sqrt{\beta_2} )\, \tr \, | S_j S_{k+1} \gamma^{(k+1)} S_j
S_{k+1}|\; .
\end{split}
\end{equation}
\end{lemma}

\subsection{Compactness of the sequence $\Gamma_{N} (t)$}

The aim of this section is to prove part i) of Theorem
\ref{thm:main}.

\begin{proof}[Proof of Theorem \ref{thm:main}, part i)]
First of all we note that the sequence $\Gamma_{N,t} = \{
\gamma^{(k)}_{N,t} \}_{k=1}^N$ is uniformly bounded in $H_-$. In
fact, from $\gamma^{(k)}_{N,t} \geq 0$ and $\tr \,
\gamma^{(k)}_{N,t} =1$ it follows immediately that
\begin{equation}\label{eq:unitball}
\| \Gamma_{N,t} \|_{H_-} = \sum_{k \geq 1} 2^{-k} \, \int \rd
\bx_k \rd \bx'_k \, |\gamma_{N,t}^{(k)} (\bx_k ; \bx'_k)|^2 \leq 1
\, .
\end{equation}
Next we prove that the sequence $\Gamma_{N,t}$ is equicontinuous
in time with respect to the metric $\rho$ (see (\ref{eq:rho}))
defined on $H_-$. To check equicontinuity we use the following
lemma, whose proof can be found in \cite{ESY} (see Lemma 9.2).

\begin{lemma}\label{lm:equi}
The sequence $\Gamma_{N, t}=\{ \gamma_{N, t}^{(k)}\}_{k = 1}^N$,
$N=1,2, \ldots$ satisfying (\ref{eq:unitball}) is equicontinuous
on $H_-$ with respect to the metric $\rho$ if and only if for
every fixed $k \geq 1$, for arbitrary $J^{(k)} \in W^{1,\infty}
(\Lambda^k \times \Lambda^k)$ and for every $\eps
>0$ there exists a $\delta > 0$ such that
\begin{equation}\label{eq:equi02}
\Big| \langle J^{(k)} , \gamma_{N,t}^{(k)} - \gamma_{N,s}^{(k)}
\rangle \Big| \leq \eps
\end{equation}
whenever $|t -s| \leq \delta$.
\end{lemma}

So, in order to prove that $\Gamma_{N,t}$ is equicontinuous, we
choose $k \geq 1$, $J^{(k)} \in W^{1,\infty}(\Lambda^k \times
\Lambda^k)$ and $\eps >0$. Then by Lemma \ref{lm:conv}, we have
\begin{equation}\label{eq:equi1}
\begin{split}
\Big| \langle J^{(k)}, &\gamma_{N,t}^{(k)} - \gamma_{N,s}^{(k)}
\rangle \Big| \leq \sum_{j=1}^k  \int_s^t \rd \tau \Big| \langle
J^{(k)}, (-\Delta_{x_j} + \Delta_{x'_j}) \gamma_{N,\tau}^{(k)}
\rangle \Big| \\ & + b \sum_{j=1}^k \Big| \int_s^t \rd \tau \int
\rd \bx_k \rd \bx'_k \rd x_{k+1} J^{(k)} (\bx_k,\bx'_k)
(\delta_{\beta} (x_j - x_{k+1}) - \delta_{\beta} (x'_j - x_{k+1}))
\\ & \hspace{3cm} \times \gamma_{N,\tau}^{(k+1)} (\bx_k,x_{k+1}, \bx'_k, x_{k+1}) \Big|
\\&+ |t-s| \left( k O(\beta^{1/2}) + k O(a^{1/2}) + k^2 O \left(
\frac{1}{Na^{1/2}}\right) \right) \, \sup_{\tau \in [0,t]} \tr
|S_1 S_2 \gamma^{(2)}_{N,\tau} S_2 S_1|\, .
\end{split}
\end{equation}
Next we note that
\begin{equation}
|\langle J^{(k)}, (-\Delta_{x_j} + \Delta_{x'_j})
\gamma_{N,\tau}^{(k)} \rangle \Big| \leq 2\| S_j^{-1} J^{(k)} S_j
\| \, \tr \, S_j \gamma_{N,\tau} S_j
\end{equation}
is uniformly bounded in $N$ and in $\tau \in [s,t]$, because of
Corollary \ref{cor:energy} and of (\ref{eq:SJ}).

As for the term on the second line of (\ref{eq:equi1}) we note
that, for fixed $\beta >0$, it is bounded by
\begin{equation}\begin{split}
b \sum_{j=1}^k \Big| \int_s^t \rd \tau \int \rd &\bx_k \rd \bx'_k
\rd x_{k+1} J^{(k)} (\bx_k,\bx'_k) (\delta_{\beta} (x_j - x_{k+1})
- \delta_{\beta} (x'_j - x_{k+1})) \gamma_{N,\tau}^{(k+1)}
(\bx_k,x_{k+1}, \bx'_k, x_{k+1}) \Big| \\ &\leq C \beta^{-3}
\sum_{j =1}^k \int_s^t \rd \tau \, \| J^{(k)} \| \, \tr \,
\gamma^{(k+1)}_{N,\tau}  \leq C_{\beta,k,J^{(k)}} |t-s|
\end{split}
\end{equation}
where we used that $J^{(k)}$ is a bounded operator (this follows
easily from the condition that its kernel lies in $W^{1,\infty}
(\Lambda^k \times \Lambda^k)$, and from the finiteness of the
volume), that the norm of $\delta_{\beta}$ is of order
$\beta^{-3}$, and that the trace of $\gamma_{N,\tau}^{(k+1)}$ is
one, for every $\tau$ and $N$. {F}rom the last three equations we
find
\begin{equation}
\Big| \langle J^{(k)}, \gamma_{N,t}^{(k)} - \gamma_{N,s}^{(k)}
\rangle \Big| \leq C \, |t-s|
\end{equation}
for a constant $C$, depending on $\beta,k,J^{(k)}$, but
independent of $N,t,s$. This implies, by Lemma \ref{lm:equi}
equicontinuity of $\Gamma_{N,t}$. The equicontinuity of
$\Gamma_{N,t}$ then implies that $\Gamma_{N,t}$ is compact in
$C([0,T], H_-)$ by the Arzela-Ascoli Theorem.
\end{proof}

\subsection{A-priori bounds on
$\Gamma_{\infty,t}$}\label{sec:apriori}

The aim of this section is to prove part ii) of Theorem
\ref{thm:main}. To this end we define a new topology in the space
of density matrices.

Denote by $\cL^1 (\cH)$ and by $\cK (\cH)$ the space of trace
class operators and, respectively, the space of compact operators
on a Hilbert space $\cH$. Moreover let $H_k = L^2 (\Lambda^k)$.
For a density matrix $\gamma^{(k)} \in \cL^1 (H_k)$, we define the
norm
\begin{equation}
\| \gamma^{(k)} \|_{\cW_k} = \tr |S_1 \dots S_k \, \gamma^{(k)} \,
S_1 \dots S_k |
\end{equation}
where $S_j = (1 -\Delta_j)^{1/2}$. We put
\begin{equation}
\cW_k = \{ \gamma^{(k)} \in \cL^1 (H_k) : \| \gamma^{(k)}
\|_{\cW_k} < \infty \}\, .
\end{equation}
We consider moreover the space
\begin{equation} \cA^{(k)} = \{
T^{(k)} = S_1 \dots S_k \, K^{(k)} \, S_1 \dots S_k : K^{(k)} \in
\cK (H_k) \}
\end{equation}
equipped with the norm
\begin{equation}
\| T^{(k)} \|_{\cA^{(k)}} = \| S_1^{-1} \dots S_k^{-1} T^{(k)}
S_1^{-1} \dots S_k^{-1} \|
\end{equation}
where $\| . \|$ denotes the operator norm. We have \[ ( \cA^{(k)}
, \| . \|_{\cA^{(k)}})^* = (\cW_k , \| . \|_{\cW_k}) .\] The
identification of $\cW_k$ as the dual space to $\cA^{(k)}$ implies
the existence of a weak star topology on $\cW_k$.



\begin{proof}[Proof of part ii) of Theorem \ref{thm:main}]
Let $\Gamma_{\infty,t} = \{ \gamma^{(k)}_{\infty,t} \}_{k \geq 1}$
be any limit point of $\Gamma_{N,t}=\{ \gamma_{N,t}^{(k)}
\}_{k=1}^N$ in the space $C([0,T],H_-)$ with respect to the metric
$\wh \rho$. By passing to a subsequence we can assume that
$\Gamma_{N,t} \to \Gamma_{\infty,t}$, for $N\to\infty$, w.r.t. the
metric $\wh \rho$. This implies that, for every fixed $t\in[0,T]$
and for every $k \geq 1$, we have
\begin{equation}\label{eq:conv1}
\gamma_{N,t}^{(k)} \to \gamma_{\infty,t}^{(k)}
\end{equation}
with respect to the weak topology of $L^2 (\Lambda^k \times
\Lambda^k)$. This follows because $\| \Gamma_{N,t}\|_{H_-} \leq 1$
and because in the unit ball, the metric $\rho$ is equivalent to
the weak$*$ topology of $H_-$. Convergence with respect to the
weak$*$ topology of $H_-$ implies then weak convergence in every
$k$-particle sector $L^2 (\Lambda^k \times \Lambda^k)$.

By Corollary \ref{cor:energy}, there exists a constant $C$ such
that
\begin{equation}
\| \gamma^{(k)}_{N,t} \|_{\cW_k} = \tr \, \Big| S_1 \dots S_k
\gamma^{(k)}_{N,t} S_k \dots S_1 \Big| \leq C^k
\end{equation}
for every $t \in [0,T]$ and $k \geq 1$. By the Banach-Alaoglu
Theorem, the sequence $\gamma^{(k)}_{N,t}$ is compact in $\cW_k$
with respect to the weak$*$ topology. In particular there exists a
subsequence $N_j \to \infty$, and $\wt \gamma^{(k)}_{\infty,t} \in
\cW_k$ such that $\gamma^{(k)}_{N_j,t} \to \wt
\gamma_{\infty,t}^{(k)}$ and \begin{equation}\label{eq:gamboun} \|
\wt \gamma^{(k)}_{\infty,t}\|_{\cW_k} = \tr \, \Big| S_1 \dots S_k
\wt \gamma^{(k)}_{\infty,t} S_k \dots S_1 \Big| \leq C^k.
\end{equation}
So, the sequence $\gamma^{(k)}_{N_j,t}$ satisfies, for $j \to
\infty$,
\begin{equation}\label{eq:conv2}
\begin{split}
\gamma_{N_j,t}^{(k)} &\to \gamma_{\infty,t}^{(k)} \quad
\text{w.r.t. the weak topology of } L^2 (\Lambda^k \times
\Lambda^k) \quad \text{and}\\
\gamma_{N_j,t}^{(k)} &\to \wt \gamma_{\infty,t}^{(k)} \quad
\text{w.r.t. the weak$*$ topology of } \cW_k\,.
\end{split}
\end{equation}
If $J^{(k)} \in L^2 (\Lambda^k \times \Lambda^k)$ then the
operator with kernel given by $J^{(k)}$ (which will be still
denoted by $J^{(k)}$) is Hilbert-Schmidt and thus compact: in
particular $J^{(k)} \in \cA_k$. Thus, using (\ref{eq:conv2}), it
is easy to verify that
\begin{equation}
\int \rd \bx_k \rd \bx'_k \, J^{(k)} (\bx_k , \bx'_k)
\gamma_{\infty,t}^{(k)} (\bx_k , \bx'_k) = \int \rd \bx_k \rd
\bx'_k \, J^{(k)} (\bx_k , \bx'_k) \wt \gamma_{\infty,t}^{(k)}
(\bx_k , \bx'_k)
\end{equation}
for every $J^{(k)} \in L^2 (\Lambda^k \times \Lambda^k)$. This
implies that $\gamma_{\infty,t}^{(k)} = \wt
\gamma_{\infty,t}^{(k)}$ as elements of $L^2 (\Lambda^k \times
\Lambda^k)$. Thus, from (\ref{eq:gamboun}), we have
\begin{equation}
\tr \, \Big|S_1 \dots S_k \gamma^{(k)}_{\infty,t} S_k \dots
S_1\Big|  \leq C^k
\end{equation}
for every $t \in [0,T]$ and $k\geq 1$. More precisely one should
say that there is a version of $\gamma^{(k)}_{\infty,t} \in L^2
(\Lambda^k \times \Lambda^k)$ which satisfies this bound (the
version we are using here is exactly the density matrix $\wt
\gamma_{\infty,t}^{(k)}$).
\end{proof}

\subsection{Non-triviality of the limit points}

\begin{proof}[Proof of part iii) of Theorem \ref{thm:main}]
Suppose $\Gamma_{\infty,t} = \{ \gamma_{\infty,t}^{(k)} \}_{k \geq
1}$ is a limit point of $\Gamma_{N,t}$ in the space $C([0,T],
H_-)$ with respect to the metric $\wh \rho$. {F}rom Section
\ref{sec:apriori}, we know that, for every fixed $t \in [0,T]$ and
$k\geq 1$, $\gamma_{\infty,t}^{(k)} \in \cW_k$ (more precisely
there is a version of $\gamma_{\infty,t}^{(k)}$ lying in the space
$\cW_k$), and that there is a subsequence $N_j \to \infty$ with
$\gamma_{N_j,t}^{(k)} \to \gamma_{\infty,t}^{(k)}$ w.r.t. the
weak$*$ topology of $\cW_k$. This means that
\begin{equation}
\tr \, J^{(k)} \left( \gamma_{N_j,t}^{(k)} -
\gamma_{\infty,t}^{(k)} \right) \to 0
\end{equation}
for $j \to \infty$, and for every $J^{(k)} \in \cA_k$ (recall that
$J^{(k)} \in \cA_k$ if and only if $S_1^{-1} \dots S_k^{-1}
J^{(k)} S_k^{-1} \dots S_1^{-1}$ is compact as operator on $L^2
(\Lambda^k)$). Next we note that, because of the finiteness of the
volume of $\Lambda$, the identity operator is an element of
$\cA_k$ (since $S_1^{-2} \dots S_k^{-2}$ is a compact operator on
$L^2 (\Lambda^k)$), and thus
\begin{equation}
\tr \, \gamma_{\infty,t}^{(k)} = \lim_{j \to \infty} \tr \,
\gamma_{N_j ,t}^{(k)} =1
\end{equation}
because $\tr \gamma_{N,t}^{(k)} =1$ for every $N,t,k$.
\end{proof}

\subsection{Convergence to the infinite BBGKY Hierarchy}

In this section we prove the last two parts of Theorem
\ref{thm:main}.

\begin{proof}[Proof of part iv) of Theorem \ref{thm:main}] Eq. (\ref{eq:limrr'})
follows from part ii) of Theorem \ref{thm:main} and from Lemma
\ref{lm:sobsob}, by the following simple argument. {F}rom Lemma
\ref{lm:sobsob} we find
\begin{equation}
\begin{split}
\Big| \int \rd {\bf x}_k \rd {\bf x}_k' \rd x_{k+1} &\rd x'_{k+1}
\, J^{(k)} (\bx_k ; \bx'_k) \, \gamma^{(k+1)}_{\infty,t} (\bx_k,
x_{k+1}; \bx'_k , x'_{k+1}) \\  \times \big( \delta_{r_1}
&(x'_{k+1} - x_{k+1}) \delta_{r'_1} (x_j -x_{k+1}) - \delta_{r_2}
(x'_{k+1} - x_{k+1}) \delta_{r'_2} (x_j - x_{k+1}) \big) \Big|
\\ \leq \; &C \big[  \|J\|_\infty + \|\nabla_j J \|_\infty \, \big]
\, (r_1 + r_2 + \sqrt{r'_1} + \sqrt{r'_2})\, \tr \, | S_j S_{k+1}
\gamma^{(k+1)}_{\infty,t} S_j S_{k+1}|\; .
\end{split}
\end{equation}
This implies, by (\ref{eq:apriori}), that the sequence
\begin{equation}
\begin{split}
\int \rd {\bf x}_k \rd {\bf x}_k' \rd x_{k+1} \rd x'_{k+1} \,
J^{(k)} (\bx_k ; &\bx'_k) \, \delta_{r} (x'_{k+1} - x_{k+1})
\delta_{r'} (x_j -x_{k+1}) \\
&\times \gamma^{(k+1)}_{\infty,t} (\bx_k, x_{k+1}; \bx'_k ,
x'_{k+1})
\end{split}
\end{equation}
has the Cauchy property for $r,r' \to 0$ and thus converges, if
$J^{(k)} \in W^{1,\infty} (\Lambda^k \times \Lambda^k)$.
\end{proof}

\begin{proof}[Proof of part v) of Theorem \ref{thm:main}]
{F}rom Lemma \ref{lm:conv} we find, for an arbitrary $J^{(k)} \in
W^{1,\infty} (\Lambda^k \times \Lambda^k)$, and for $N$ large
enough,
\begin{equation}\label{eq:conver1}
\begin{split}
\langle J^{(k)} , &\gamma^{(k)}_{N,t} \rangle = \langle J^{(k)} ,
\gamma^{(k)}_{N,0} \rangle -i \sum_{j=1}^k \int_0^t \rd s \int \rd
\bx_k \rd \bx'_k J^{(k)} (\bx_k,\bx'_k) (-\Delta_{x_j} +
\Delta_{x'_j}) \gamma_{N,s}^{(k)} (\bx_k,\bx'_k) \\ & -ib
\sum_{j=1}^k \int_0^t \rd s \int \rd \bx_k \rd \bx'_k \rd x_{k+1}
J^{(k)} (\bx_k,\bx'_k) (\delta_{\beta} (x_j - x_{k+1}) -
\delta_{\beta} (x'_j - x_{k+1})) \\
&\hspace{3cm} \times \gamma_{N,s}^{(k+1)} (\bx_k,
x_{k+1},\bx'_k,x_{k+1})\\&+ t O(\beta^{1/2}) + t o (1)
\end{split}
\end{equation}
where $o(1) \to 0$  for $N \to \infty$. By passing to a
subsequence we can assume that $\Gamma_{N,t} \to \Gamma_{\infty,t}
= \{ \gamma_{\infty,t}^{(k)} \}_{k \geq 1} \in C([0,T], H_-)$
w.r.t. the metric $\wh \rho$. Since $\| \Gamma_{N,t} \|_{H_-} \leq
1$, and since the metric $\rho$ on the unit ball of $H_-$ is
equivalent to the weak$*$ topology, it follows that
$\gamma^{(k)}_{N,t} \to \gamma^{(k)}_{\infty,t}$ w.r.t. the weak
topology of $L^2 (\Lambda^k \times \Lambda^k)$, for every fixed $k
\geq 1$ and $t \in [0,T]$. For $J^{(k)} \in W^{2,\infty}
(\Lambda^k \times \Lambda^k)$, we also have $J^{(k)} \in L^2
(\Lambda^k \times \Lambda^k)$ (because of the finiteness of the
volume). Hence
\begin{equation}\label{eq:conver2}
\langle J^{(k)}, \gamma_{N,t}^{(k)} - \gamma_{\infty,t}^{(k)}
\rangle \to 0 \quad \text{and} \quad \langle J^{(k)},
\gamma_{N,0}^{(k)} - \gamma_{\infty,0}^{(k)} \rangle \to 0
\end{equation}
for $N \to \infty$.

As for the second term on the r.h.s. of (\ref{eq:conver1}) we note
that, from $J^{(k)} \in W^{2,\infty} (\Lambda^k \times
\Lambda^k)$, it also follows that $\Delta_{x_j} J^{(k)} (\bx_k;
\bx'_k)$ and $\Delta_{x'_j} J^{(k)} (\bx_k; \bx'_k)$ are elements
of $L^2 (\Lambda^k \times \Lambda^k)$. This implies that
\begin{equation}
\sum_{j=1}^k \int \rd \bx_k \rd \bx'_k \, \Delta_j J^{(k)} (\bx_k
; \bx'_k ) \left( \gamma_{N,s}^{(k)} (\bx_k ; \bx'_k) -
\gamma_{\infty ,s}^{(k)} (\bx_k ; \bx'_k) \right) \to 0
\end{equation}
for $N \to \infty$, and for every $s \in [0,t]$. By Lebesgue
Theorem on the dominated convergence, we find
\begin{equation}\label{eq:conver3}
\sum_{j=1}^k \int_0^t \rd s \int \rd \bx_k \rd \bx'_k J^{(k)}
(\bx_k ;\bx'_k) (-\Delta_{x_j} + \Delta_{x'_j})
\left(\gamma_{N,s}^{(k)} (\bx_k ;\bx'_k) - \gamma_{\infty,s}^{(k)}
(\bx_k ; \bx'_k) \right)\to 0
\end{equation}
for $N \to \infty$ and for every fixed $t \in [0,T]$.

Finally we consider the limit $N \to \infty$ of the last term on
the r.h.s. of (\ref{eq:conver1}). {F}rom Lemma \ref{lm:sobsob}, we
have
\begin{multline}
\int \rd \bx_k \rd \bx'_k \rd x_{k+1} \, J^{(k)} (\bx_k ; \bx'_k)
\left(\delta_{\beta} (x_j - x_{k+1}) - \delta_{\beta} (x'_j -
x_{k+1}) \right) \gamma_{N,s}^{(k+1)} (\bx_k , x_{k+1} ; \bx'_k,
x_{k+1} ) \\ = \int \rd \bx_k \rd \bx'_k \rd x_{k+1}\rd x'_{k+1}
\, J^{(k)} (\bx_k ; \bx'_k) \left(\delta_{\beta} (x_j - x_{k+1}) -
\delta_{\beta} (x'_j -x_{k+1}) \right) \\ \times \delta_{\eta}
(x_{k+1} - x'_{k+1}) \gamma_{N,s}^{(k+1)} (\bx_k , x_{k+1} ;
\bx'_k, x'_{k+1}) + O (\eta)
\end{multline}
where $O(\eta)$ is independent of $\beta, N$ and $s$. At this
point we can take the limit $N \to \infty$ with fixed $\beta$ and
$\eta$. Since $J^{(k)} \in W^{2,\infty} (\Lambda^k \times
\Lambda^k)$, it is easy to check that, for fixed $\beta , \eta
>0$, $J^{(k)} (\bx_k ; \bx'_k) \delta_{\beta} (x_j -x_{k+1})
\delta_{\eta} (x_{k+1} -x'_{k+1})$ is an element of $L^2
(\Lambda^{k+1} \times \Lambda^{k+1})$. Hence
\begin{equation}\label{eq:conver4}
\begin{split}
\int \rd \bx_k \rd \bx'_k \rd x_{k+1}\rd x'_{k+1} \, &J^{(k)}
(\bx_k ; \bx'_k) \left(\delta_{\beta} (x_j - x_{k+1}) -
\delta_{\beta} (x'_j -x_{k+1}) \right) \delta_{\eta} (x_{k+1} -
x'_{k+1}) \\ &\times \left( \gamma_{N,s}^{(k+1)} (\bx_k , x_{k+1}
; \bx'_k, x'_{k+1}) - \gamma_{\infty,s}^{(k+1)} (\bx_k, x_{k+1} ;
\bx'_k, x'_{k+1}) \right) \to 0
\end{split}
\end{equation}
for $N \to \infty$, uniformly in $s$. Using (\ref{eq:conver2}),
(\ref{eq:conver3}), and (\ref{eq:conver4}), it follows from
(\ref{eq:conver1}), that
\begin{equation}
\begin{split}
\langle J^{(k)} , \gamma_{\infty,t}^{(k)} \rangle = \; &\langle
J^{(k)}, \gamma_{\infty,0}^{(k)} \rangle -i \sum_{j=1}^k \int_0^t
\rd s \, \int \rd {\bf x}_k \rd {\bf x}_k' \, J^{(k)} ({\bf x}_k;
{\bf x}_k') \, ( -\Delta_j + \Delta_j') \gamma^{(k)}_{\infty,s}
({\bf x}_k; {\bf x}_k') \\
&-ib \sum_{j=1}^k \int_0^t \rd s \int \rd {\bf x}_k \rd {\bf x}_k'
\rd x_{k+1} \rd x'_{k+1} \, J^{(k)} ({\bf x}_k ;{\bf x}_k')
(\delta_{\beta} (x_j - x_{k+1}) - \delta_{\beta} (x_j' - x_{k+1}))
\\ &\times \delta_{\eta} (x_{k+1} -x'_{k+1})
\gamma^{(k+1)}_{\infty,s} (\bx_{k}, x_{k+1}; \bx'_{k}, x_{k+1}') +
O(\beta^{1/2}) + O (\eta)
\end{split}
\end{equation}
for any fixed $t$ and $k$. Finally, we apply Lemma \ref{lm:sobsob}
to replace $\delta_{\eta} (x_{k+1} -x'_{k+1})$ by $\delta (x_{k+1}
-x_{k+1}')$ and $\delta_{\beta} (x_j - x_{k+1})$ (respectively,
$\delta_{\beta} (x'_j - x_{k+1})$) by $\delta (x_j - x_{k+1})$
(respectively, by $\delta (x'_j - x_{k+1})$). The error here is of
order $\beta^{1/2} + \eta$. Hence, letting $\eta \to 0$ and $\beta
\to 0$ we find
\begin{equation}
\begin{split}
\langle J^{(k)} , \gamma_{\infty,t}^{(k)} \rangle = \; &\langle
J^{(k)}, \gamma_{\infty,0}^{(k)} \rangle -i \sum_{j=1}^k \int_0^t
\rd s \, \int \rd {\bf x}_k \rd {\bf x}_k' \, J^{(k)} ({\bf x}_k ;
{\bf x}_k') \, ( -\Delta_j + \Delta_j') \gamma^{(k)}_{\infty,s}
({\bf x}_k ; {\bf x}_k') \\
&- ib \sum_{j=1}^k \int_0^t \rd s \int \rd {\bf x}_k \rd {\bf
x}_k' \rd x_{k+1} \, J^{(k)} ({\bf x}_k ; {\bf x}_k') (\delta (x_j
- x_{k+1}) - \delta (x_j' - x_{k+1}))
\\ &\times
\gamma^{(k+1)}_{\infty,s} (x_1, \dots, x_{k+1}; x_1',\dots,
x_{k}', x_{k+1})\; .
\end{split}
\end{equation}
\end{proof}

\appendix

\section{Sobolev Type Inequalities}\label{app:sob}

\begin{lemma}\label{lm:sob}
\begin{itemize}
\item[i)] Suppose $V \in L^{3/2} (\Lambda)$, and $\psi \in W^{1,2}
(\Lambda)$. Then
\begin{equation}
\int \rd x |\psi (x)|^2 \frac{1}{a^2} V(x/a) \leq C \|
V\|_{L^{3/2}(\Lambda)} \, \left( \| \nabla \psi \|^2 + \| \psi
\|^2 \right)^{1/2} \, .
\end{equation}
\item[ii)] Suppose $V \in L^1 (\Lambda)$. Then, considering
$V(x-y)$ as an operator on $L^2 (\Lambda , \rd y) \otimes L^2
(\Lambda , \rd x)$ we have the operator inequality
\begin{equation}
\frac{1}{a^3} V\left(\frac{x-y}{a}\right) \leq  C \| V \|_{L^1} \,
(1 - \Delta_x) ( 1-\Delta_y) \, .
\end{equation}
\end{itemize}
\end{lemma}
\begin{proof} The proof of i) can be
found in \cite{ESY}. The proof of ii) is in \cite{EY}.
\end{proof}

\end{document}